\documentclass[%
preprint,
 amsmath,amssymb,
prl,
]{revtex4-1}
\usepackage{graphicx}% Include figure files
\usepackage{dcolumn}% Align table columns on decimal point
\usepackage{bm}% bold math
\usepackage{color}
\usepackage{epstopdf}
\usepackage{gensymb}
\usepackage[english]{babel}
\usepackage{float}
\usepackage{xr}
\usepackage{sidecap}
\usepackage{mathrsfs,amsmath}

\usepackage[normalem]{ulem}

\begin{document}

\title{Multi-functional resonant acoustic \\ networks for wave routing}

\author{Olivier Richoux}
\affiliation{%
LAUM, UMR-CNRS 6613, Le Mans University, Av. O. Messiaen, 72085 Le Mans, France}%

\author{Vassos Achilleos}
\affiliation{%
LAUM, UMR-CNRS 6613, Le Mans University, Av. O. Messiaen, 72085 Le Mans, France}%

\author{Georgios Theocharis}
\affiliation{%
LAUM, UMR-CNRS 6613, Le Mans University, Av. O. Messiaen, 72085 Le Mans, France}%

\author{Ioannis Brouzos}
\affiliation{%
Department of Physics, University of Athens, 15771 Athens, Greece}
\author{Fotios Diakonos}
\affiliation{%
Department of Physics, University of Athens, 15771 Athens, Greece}

\date{\today}

\begin{abstract}
We demonstrate that asymmetric 3-port devices can be used to design a multi-functional set-up operating as a symmetric combiner and splitter at the same frequency.  For a concrete implementation of the proposed multi-functionality protocol, we employ a resonant acoustic 3-port device operating in a subwavelength regime, showing splitting/combining abilities with nearly perfect transmission in presence of losses, induced by the interactions of the system's resonances. Finally, as an extension of the 3-port multi-functionality concept, we propose a new type of 4-port networks, consisting in an assembly of 3-port unit cells with combining/splitting properties, capable to achieve various wave guiding features with perfect transmission depending on the phase of the inputs.
\end{abstract}

\maketitle

Wave routing has always been a great challenge for many physical applications.
Numerous works focus on the design of multi-port devices capable to guide waves from specific inlets to different outlets. 
Such devices include power divider or combiner (which divides or combines one or several inputs into several or one ports) or multiplexer systems which selects one of several inputs and transfers this input into a specific port. For example, in microwave engineering, guiding devices such as power dividers, circulators, filters couplers and multiplexer are commonly used \cite{Pozar}. 
Multiport devices can be realized in different ways employing resonant structures \cite{Rosenberg,Pozar,Mouadili},
photonic crystals \cite{Yu,Park,Shinya}, 
metamaterials \cite{Tseng,Ourir2} 
as well as waveguides with multimode interferences \cite{Han,Sun}. In electromagnetics, selected filters and multiplexers have also been implemented by using structured materials as nonuniform waveguide network \cite{Feigenbaum2010a,Feigenbaum2010b} and 
complex photonic circuits are studied and designed for optical signal processing or computing in integrated optics~\cite{Parinaz}. 

In acoustics, two-port systems have been studied and different phenomena including Coherent Perfect Absorption\cite{Merkel,Romero}, reflectionless transmission \cite{Merkel2018} or perfect transparency \cite{Tao} have been revealed. Nevertheless, it appears that only a few works are dealing with systems with more than two ports. As example, three-port systems based on resonant scatterers, have been proposed to achieve Coherent Perfect Absortion/Coherent Perfect Transmission \cite{Richoux2018} or channeling \cite{Zhang}, by tuning the geometry of the system and the wave input. PT-symmetric three-port systems using spatiotemporal modulation or circulating fluid, have been employed to obtain non reciprocal acoustic circulator \cite{Fleury2015,Fleury2014}. Directional excitation, asymmetric reflection and wave router for multi-ports have been developed using metamaterials and/or PT-symmetric systems \cite{Fu,ZhangAc,Zhu,Dai,Ramezani}. Finally, a splitter system has been realized using acoustic topological network \cite{He}.

In this work, we demonstrate (both theoretically and numerically) that an acoustic resonant subwavelength three-port can be used to obtain a symmetric splitter and a symmetric combiner with a quasi perfect transmission despite the presence of losses in the system. The device is composed of identical connected waveguides side-loaded by resonant scatterers. The control of the wave is made possible using induced transparency resonance \cite{Mouadili} coming from the interaction of the resonances of individual scatterers through the waveguide. Tuning these resonances, one can achieve wave guiding with almost perfect transmission. 
By arranging several 3-port systems with these specific properties, we propose an optimally designed network which allows for multiple functionalities illustrated here by a 4-port case. We design a directional coupler with prescribed wave guiding abilities depending on the relative phase of the input waves. We demonstrate using 3D simulation the feasibility of this device.

\begin{figure}
\centering
\includegraphics[width=9cm]{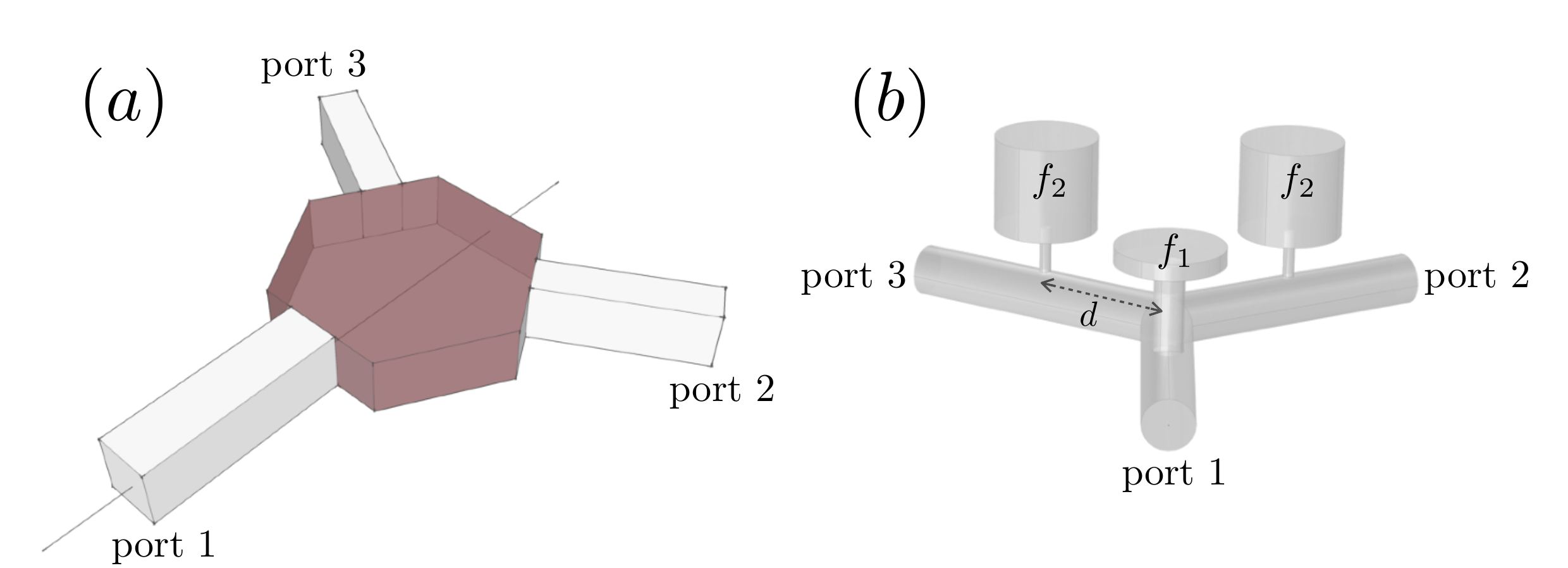}
\caption{\label{fig_fig1} (a) Sketch of a mirror symmetric 3-port system. The line denotes the axis of the symmetry (b) Schematic of the 3-port acoustic device made of 3 connected waveguides side loaded by 3 HRs with resonance frequency $f_1$ and $f_2$. The distance between each HR and the central connection is $d$.} 
\end{figure}

Let us consider a system connected to three identical input/output ports as shown in Fig. \ref{fig_fig1}(a), exhibiting mirror symmetry with respect to one port \textit{e.g} port 1 in Fig. \ref{fig_fig1}(a).

Assuming single mode propagation in each channel, we may write the corresponding scattering matrix as
\begin{equation} 
\label{Eq-Scattering}
\begin{pmatrix}
p_1^- \\
p_2^- \\
p_3^-
\end{pmatrix}
=\begin{pmatrix}
r & t & t\\
t & r' & t'  \\
t & t' & r' \\
\end{pmatrix}
\begin{pmatrix}
p_1^+ \\
p_2^+ \\
p_3^+
\end{pmatrix} 
= S
\begin{pmatrix}
p_1^+ \\
p_2^+ \\
p_3^+
\end{pmatrix}.
\end{equation}
$r$ and $t$ are the complex, frequency dependent, reflection and transmission coefficients  from port 1 while $r'$ and $t'$ are the reflection and transmission coefficients from port 2 or 3. In Eq.~(\ref{Eq-Scattering}) the vectors  $(p_1^+,p_2^+,p_3^+)^T$ and
$(p_1^-,p_2^-,p_3^-)^T$
describe the incoming and outgoing waves respectively, while the numbering denotes the corresponding port. 
Tuning the reflection and transmission coefficients allows us to 
achieve a system with prescribed functionalities at an operating frequency $f^*$.
Here we are interested in designing a 3-port system that acts simultaneously as a symmetric splitter and a symmetric combiner with perfect transmission at this operating frequency $f^*$.\\
\textit{Conditions for Symmetric Splitter:} We consider an input from port 1. We want this to be equally transmitted to ports 2 and 3 without loss at the frequency $f^*$. We thus first require that there is no reflection, namely:
\begin{equation}
r(f^*)=0.
\label{Eq:eq1}
\end{equation} 
At the same time, we require perfect transmission at the same frequency $f^*$. For this, we demand the scattering matrix
to be unitary, satisfying the following relation $\bar{S}(f^*)*S(f^*) = I$ (where $\bar{S}$ is the conjugate transpose of $S$ and $I$ is the identity matrix) which further constraints its elements as follows
\begin{align}
& r'(f^*)+t'(f^*) = 0, \label{Eq:eq2}\\
& \vert t(f^*) \vert = 1/\sqrt{2} \label{Eq:eq3}.
\end{align}
Thus, Eqs. (\ref{Eq:eq1}-\ref{Eq:eq3}) present the conditions to achieve a symmetric splitter from port 1 with a perfect transmission (no loss).\\
\textit{Conditions for Symmetric Combiner:} The symmetric combiner is simultaneously achieved from the same above conditions. Indeed, Eq. (\ref{Eq-Scattering}) combined with Eq. (\ref{Eq:eq2}) ensures that an input defined by $p_2^+ = p_3^+$ and $p_1^+=0$ is only guided to the remaining channel since we obtain $p_2^- = p_3^- = 0$ (it is a consequence of the time reversal symmetry) at the frequency $f^*$. In the same time, the perfect transmission is ensured by the Eq. (\ref{Eq:eq3}). As a consequence of these conditions, we directly obtain $\vert r'(f^*) \vert = \vert t'(f^*)\vert = 1/2$. It is also interesting to remark that by imposing conditions for symmetric splitter with perfect transmission, we obtain directly the symmetric combiner property. Note that the proposed constraints could be achieved in system supporting different kinds of waves \textit{e.g.} optical, electrical, acoustical or quantum. 

In this work, we implement such a device using a 3-port acoustic structure made of three identical waveguides connected by a symmetric Y-connection [see Fig. \ref{fig_fig1}(b)]. Each waveguide is side-loaded by a Helmholtz resonator (HR) at a distance $d$ from the Y-connection. The HRs are characterized by their resonance frequencies and their inherent losses (viscous and thermal effects at the boundaries) depending on their geometry. The symmetry of the system can be controlled by tuning the resonance frequencies of the HRs. Here we choose that the first waveguide is side-loaded by a HR with resonance frequency $f_1$ while the other two by a HR with a resonance $f_2$.

For sufficiently low frequencies (below the first cut-off frequency of the waveguide where only the plane mode is propagative)
 the acoustic device is characterised by a scattering matrix of the form of Eq. \ref{Eq-Scattering}. The coefficients $r$, $r'$, $t$ and $t'$ are analytically determined using the Transfer Matrix Method (TMM) where the visco-thermal losses in the waveguides and in the resonators are taken into account \cite{Richoux2018}. These coefficients depend both on frequency and on the geometrical characteristics of the structure \textit{e.g.} the distance $d$, necks and cavities of the HRs, etc. Below we use these analytical expressions in order to obtain the proper configuration whose coefficients satisfy Eqs. (\ref{Eq:eq1})-(\ref{Eq:eq3}), by
performing a numerical optimization based on simplex derivative free method Nelder-Mead \cite{opt} at a prescribed operating frequency of $f^*=200$ Hz.

\begin{figure}
\includegraphics[width=9cm]{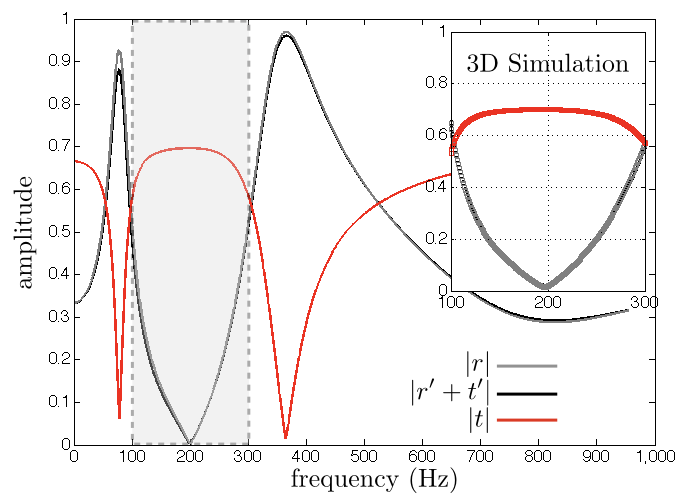}
\caption{\label{fig_RandT_TMM} Amplitude of $r$ (gray line), $t$ (red line) and $r'+t'$ (black line) determined using the TMM method with the optimized device showing splitting/combining properties with perfect transmission at $f^*=200$ Hz. The inset presents the results using a 3D FEM simulation for $f \in [100:300]$ Hz showing splitting/combining properties with perfect transmission at $f_s^*=197$ Hz.}
\end{figure}

The optimized geometrical parameters are given in \cite{opto} achieving a convergence up to 97\%. The analytical expressions of the coefficients are presented in Fig. \ref{fig_RandT_TMM} where the gray line corresponds to $\vert r \vert$, the black line to $\vert r' + t' \vert$ and the red one to $\vert t \vert$  satisfying Eqs. (\ref{Eq:eq1}-\ref{Eq:eq3}) at the operating frequency $f^*=200$ Hz.  
Indeed the analytical results give $\vert r(f^*) \vert = 1.3 \times10^{-3}$ and $ \vert t(f^*) \vert  = 0.696 \approx 1/\sqrt{2}$ and consequently $\vert r'(f^*) \vert= 0.496 \approx 1/2$ and $\vert t'(f^*) \vert = 0.494\approx 1/2$. This optimized 3-port device is achieved through the induced transparency resonance \cite{Mouadili}, localized between two transmission zeros, coming from the interaction of the detuned resonances of each HR. These resonances correspond to the two peaks in $\vert r \vert$ and $\vert r' + t' \vert$ (or dips in $\vert t \vert$) at $f_1 = 363$ Hz and $f_2 = 78 $ Hz surrounding the operating frequency.

To verify the above analytical results obtained using the TMM method, we compared them with 3D FEM simulation with visco-thermal losses at the boundaries of the system (walls of the waveguides and resonators). 
The results of the simulation are
shown in the inset of Fig. \ref{fig_RandT_TMM} for a frequency range from $100$ Hz to $300$ Hz as highlighted by the gray area.
The operating frequency is found at $f_s^*=197$ Hz very close to the analytical one $f^*=200$ Hz.
This small disagreement is essentially due to the 1D assumptions (monomode
propagation in the waveguide and no evanescent waves
around the scatterers) of the analytical model.

\begin{figure}
\centering
\includegraphics[width=6cm]{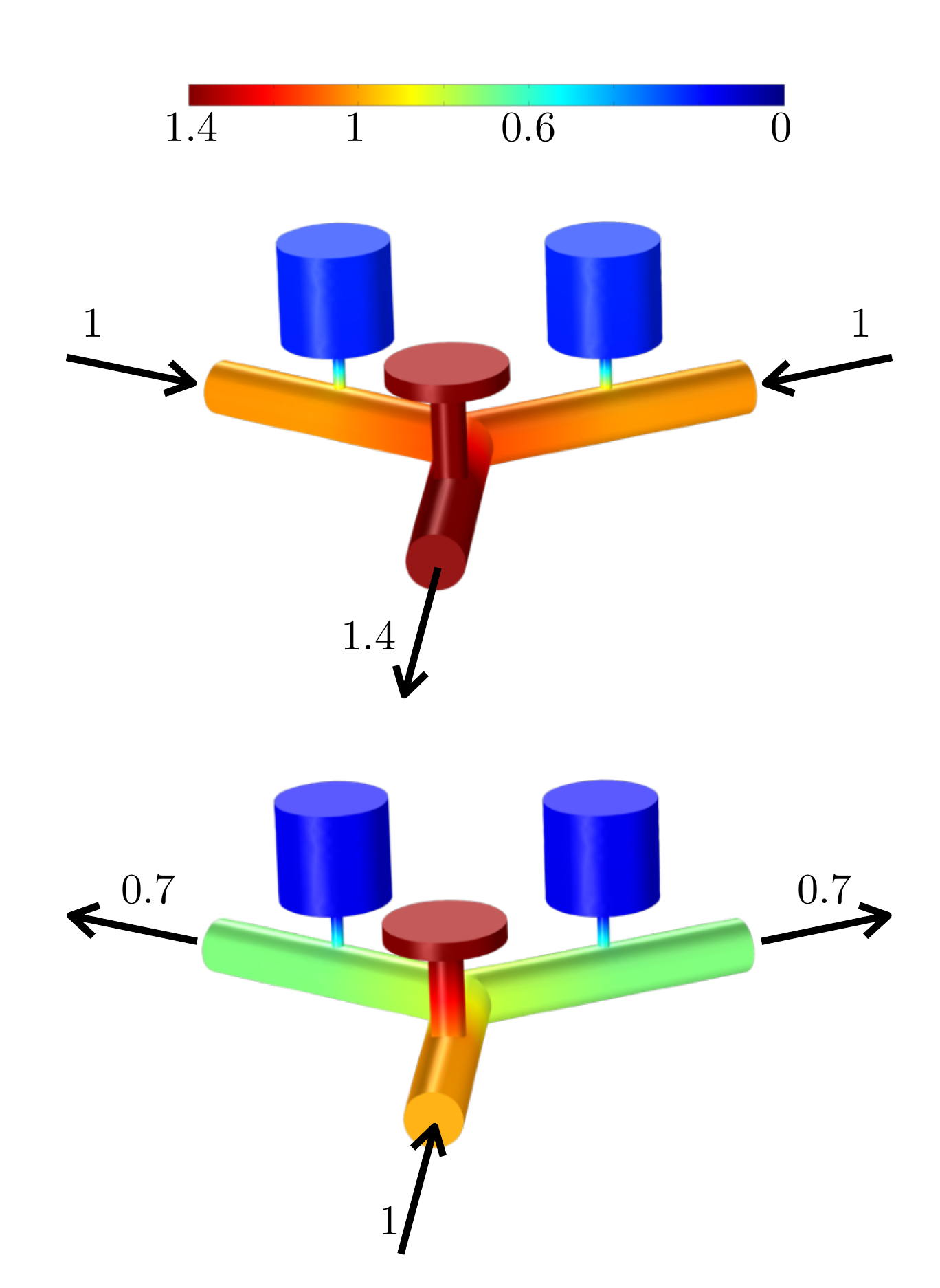}
\caption{\label{fig_field} Simulated acoustic field in the optimized 3-port device for $f_s^*=197$ Hz .(a) Combiner configuration with $p_2^+=p_3^+=1$ and $p_1^+ =0$ giving $\vert p_1^- \vert = 1.4 \approx \sqrt{2}$ and  $\vert p_2^- \vert =\vert p_3^-\vert \approx 0$. (b) Splitter configuration with $p_1^+ =1$ and $p_2^+=p_3^+=0$ giving $p_1^- \approx 0$ and $\vert p_2^-\vert=\vert p_3^-\vert = 0.7 \approx 1/ \sqrt{2}$.}
\end{figure}

In addition the HRs are considered as point scatterers in the model whereas in the FEM simulation, their sizes are small but finite. For illustration purposes, using the data from the simulation, we show both the splitting and the combining operation in Figs. \ref{fig_field}(a) and \ref{fig_field}(b) respectively at the operating frequency $f_s^*=197$ Hz. When $p_2^+=p_3^+=1$ [see Fig. \ref{fig_field}(a)], we find that $\vert p_1^- \vert = 1.4 \approx 2\vert t \vert = \sqrt{2}$ and $p_2^-=p_3^- \approx 0$ validating that the system acts as a symmetric combiner with almost perfect transmission. Furthermore when only the port $1$ is illuminated [see Fig. \ref{fig_field}(b)] with $p_1^+=1$, the wave is guided to ports 2 and 3 with an amplitude given by $\vert p_2^-\vert=\vert p_3^-\vert = 0.7 \approx \vert t \vert = 1/ \sqrt{2}$ possessing splitter property. Due to the fact that we choose HRs as scatterers, the operations of the optimized 3-port occur in a subwavelength regime. More precisely, for the acoustic set-up described here $\lambda^*/(2 d) \approx 6$ where $\lambda^*$ is the wavelength at the operating frequency.  

\begin{figure}
\includegraphics[width=6cm]{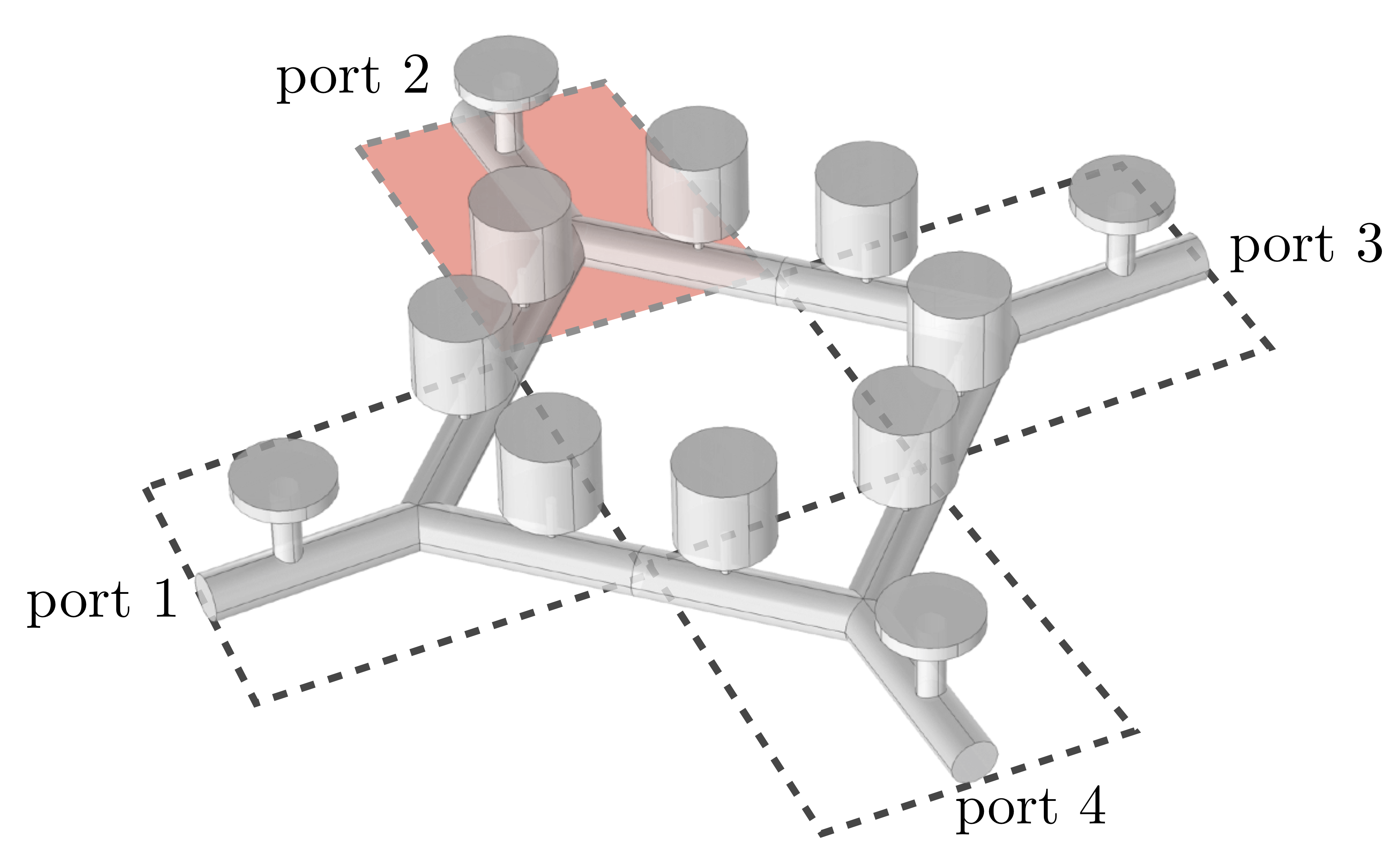}
\caption{\label{4ports_device} 4-port acoustic device constituted by 4 building blocks corresponding to identical optimized 3-port systems (marked by the dashed lines and highlighted by the color box).}
\end{figure}

\begin{figure}
\includegraphics[width=8cm]{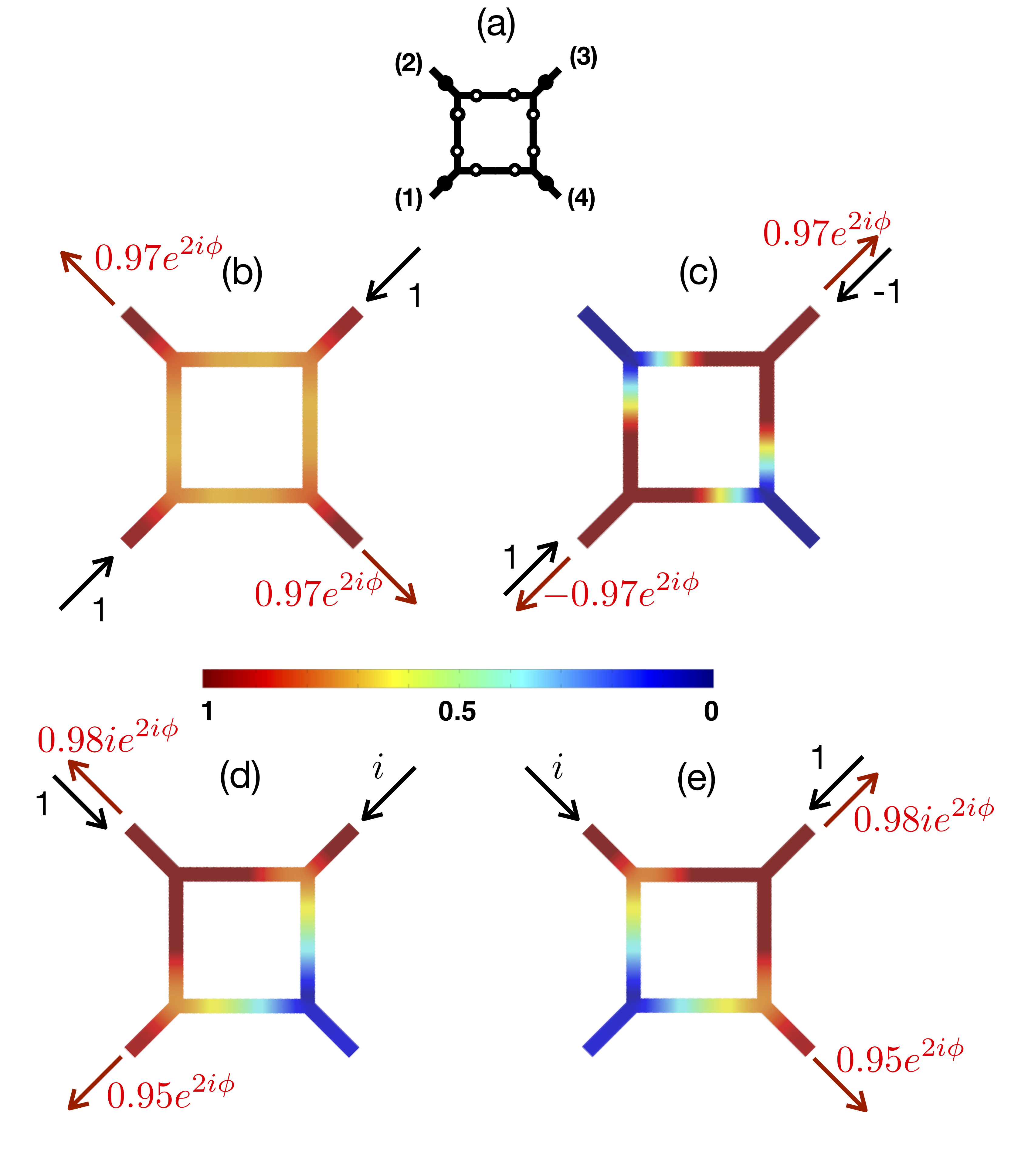}
\caption{\label{network} (a) Schematic of the 4-port network. The ports are numbered. The black and white circles denote the HRs with resonance frequency $f_1$ and with resonance frequency $f_2$ respectively. (b) Field amplitude in the 4-port device when the inputs are defined by $(1,0,1,0)^T$. (c) Field amplitude in the 4-port device when the inputs are defined by $(1,0,-1,0)^T$. (d) Field amplitude in the 4-port device when the inputs are defined by $(0,1,i,0)^T$. (e) Field amplitude in the 4-port device when the inputs are defined by $(0,i,1,0)^T$. All the fields are the results of a 3D FEM simulation. The value of the phase $2\phi$ is given by $2\phi \approx 0.42\pi$. For all the cases, the operating frequency is $f_s^*=197$ Hz. The 4-port device is built with optimized 3-port systems where we remove $5\times10^{-2}$ m of tube in the external branches to tune the transmission $t'$ and reflexion $r'$ coefficients. The colormap denotes the amplitude pressure field and the arrows gives the pressure waves.}
\end{figure}

The great advantage of this device, with the same waveguide's section at each port (as opposed to other alternatives \textit{i.e.} \cite{Zhang}) is that we can easily design a plethora of multi-port network configurations by connecting 3-port blocks.
Using the simple characteristics of this elementary cell, we can assigne specific properties to the scattering matrix of the network by choosing the orientation of each block. To illustrate this, among the numerous possible configurations, here we study a network composed by 4 identical blocks as illustrated in the Fig. \ref{4ports_device}. The resulting 4-port device features C4 symmetry due to the chosen orientation of the building blocks. 
We arrange the unit-cells such as the HRs with resonance $f_1$ are located on the external branches while the rest HRs with resonance $f_2$ are in the internal ring depicted in Fig. \ref{network}(a) with black and white dots respectively. 
Below we illustrate that such a network may acquire different routing operations at the prescribed frequency $f^*$ selected
by the input configuration.

Using the star product, the scattering matrix of this 4-port network is determined by the elements of the 3-port scattering matrix given in Eq. \ref{Eq-Scattering} as we show in detail in the supplementary material. Since we assume that our 3-port building blocks act as splitters/combiners with a nearly perfect transmission, Eqs. (\ref{Eq:eq1}-\ref{Eq:eq3}) are satisfied, and the 4-port scattering matrix can be simplified to the following form at frequency $f^*$
\begin{equation} 
\label{Eq-Scattering5}
S_{\mbox{4x4}}^{(1)}
=\frac{1}{2}e^{2i\phi}\begin{pmatrix}
i & 1 & -i & 1 \\
1 & i &  1 & -i \\
-i &  1 & i &  1 \\
 1 & -i & 1 & i \\
\end{pmatrix},
\end{equation}
where $\phi$ is the phase of $t$. To obtain Eq. (\ref{Eq-Scattering5}) we add an additional condition on the phase of $t'$ (see supplementary material) which is easily achieved by changing the waveguide length of the 3-port building blocks.
This network, satisfying Eq. (\ref{Eq-Scattering5}), owing to the chosen orientation of the building blocks, is able to perform four different kinds of routing operation  with an almost perfect transmission at the same frequency $f^*$ solely by changing the input vector.

The first two operations concern the scattering from two diagonal ports with waves of the same amplitude but differing in phase.
When the phase of the two inputs is equal, the device can route waves from two diagonal ports to the other two acting as an orthogonal directional coupler.
 To illustrate this functionality at the prescribed frequency $f_s^*$, we perform 3D FEM simulation with an input of the form $\vert \psi^+\rangle \equiv (p_1^+,p_2^+,p_3^+,p_4^+)^T=(1,0,1,0)^T$. As shown in Fig. \ref{network}(b), the wave is equally guided towards the ports 2 and 4 with an almost perfect transmission since we find $ p_2^-= p_4^-= 0.97e^{i 0.42 \pi}$ and $\vert p_1^-\vert = \vert p_3^-\vert = 0.03$. On the other hand, when diagonal ports are excited by waves of the same amplitude but opposite phase $\vert \psi^+\rangle=(1,0,-1,0)^T$ then the network becomes transparent, \textit{i.e.} the remaining ports are muted. This is also confirmed by the simulations results shown in Fig. \ref{network}(c) with $ p_1^- = -0.97 e^{i 0.42 \pi}$, $ p_3^- = 0.97 e^{i 0.42 \pi}$ and $ \vert p_2^-\vert=\vert p_4^-\vert= 0.03$. Note that, as indicated in Figs. \ref{network}(b)-(c), the relative phase between the outputs is the same as for the inputs. 

The additional operations of the network satisfying Eq. \ref{Eq-Scattering5} concern the scattering from two consecutive ports with waves of equal amplitude and varying phase. 
For example when the input is defined by $\vert \psi^+\rangle = (0,1,i,0)^T$, as shown in Fig. \ref{network}(d), the input waves from ports 2 and 3 are guided to ports 1 and 2 leaving the port 4 muted. Thus the device acts as a \textcolor{red}{counter-clockwise directional coupler} with an almost perfect transmission as found by the numerical results achieving $p_1^-= 0.95 e^{i 0.42 \pi}$ and $p_2^-= 0.98 i e^{i 0.43 \pi}$.
When the inputs are inverted, \textit{i.e.} $\vert \psi^+\rangle =(0,i,1,0)^T$, the operation of the network is also reversed and it acts as a clockwise directional coupler. This functionality is illustrated in Fig. \ref{network}(e). Once again, the relative phase of the inputs is inherited to the outputs. Note that, due to the symmetry of the system, these functionality holds for any two consecutive ports.

This work proposes the design of complex networks using subwavelength splitter/combiner 3-port systems as the network building blocks. 
Thus, we are able to manipulate the coefficients of the scattering matrix of a complex network to obtain prescribed wave routing by playing with orientation and characteristics of the 3-port unit-cell.
Aiming at particular wave routing operations, we construct a compact 4-port network by connecting four of the aforementioned building blocks in a symmetric orientation. 
The resulting network is capable of guiding incident waves to different outgoing channels, \textit{i.e.} orthogonal, clockwise and counter-clockwise directional couplers, selected only by the relative phase of the input. 
Alternative input/output functionalities, could also be achieved by connecting the building blocks with different orientations.
This concept of designing networks with multi-functionalities can also be applied for wave routing or filtering in different domains of physics as acoustics, electromagnetisms, optics or microwaves.

This work is supported by “Le Mans Acoustique” and funded by the “Pays de la Loire” French Region and the “European Regional Development Fund”. I.B. research is financed by the project CS.MICRO funded under the program "Etoiles Montantes" of the “Pays de la Loire” French Region.

%%%%%%%%%%%%%%% Bibliography 


\begin{thebibliography}{99}
\bibitem{opto}
The dimensions of HR1 are: length of the neck $\ell_{n1}=8.75\times10^{-2}$ m, radius of the neck $r_{n1}=1.49\times10^{-2}$ m, length of the cavity $\ell_{c1}=1.63\times10^{-2}$ m and radius of the cavity $r_{c1}=5.31\times10^{-2}$ m. The dimensions of HR2 are: length of the neck $\ell_{n2}=4.86\times10^{-2}$ m, radius of the neck $r_{n2}=0.55\times10^{-2}$ m, length of the cavity $\ell_{c2}=9.38\times10^{-2}$ m and radius of the cavity $r_{c2}=5.11\times10^{-2}$ m. The radius of the main waveguides is $r_t=2.5\times10^{-2}$ m. The distance between each HR and the central connection is $d = \ell/2$ with $\ell = 27;76\times 10^{-2}$ m the length of each waveguide.
\end{thebibliography}
\end{document}